\documentclass[oldversion]{aa}

\usepackage{epsfig}
\usepackage{mdwlist}
\usepackage[switch]{lineno}
\usepackage{graphicx}
\usepackage{txfonts}
\usepackage{natbib}

\bibpunct{(}{)}{;}{a}{}{,} 

\def\ecs{erg~cm$^{-2}$s$^{-1}$}
\def\lum{erg~s$^{-1}$}
\def\bron{2S 0918-549}

\begin{document}

\title{Achromatic late-time variability in thermonuclear X-ray bursts}
\subtitle{An accretion disk disrupted by a nova-like shell?}

\titlerunning{Achromatic late-time variability in thermonuclear X-ray bursts}
\authorrunning{in 't Zand, Galloway \& Ballantyne}

\author{J. J. M.~in~'t~Zand\inst{1}, D.K. Galloway\inst{2} \& D.R. Ballantyne\inst{3}}

\institute{     SRON Netherlands Institute for Space Research, Sorbonnelaan 2,
                3584 CA Utrecht, the Netherlands; {\tt jeanz@sron.nl}
           \and
                Center for Stellar and Planetary Astrophysics \&
                School of Physics, Monash University, VIC 3800, Australia
           \and
                Center for Relativistic Astrophysics, School of Physics,
                Georgia Institute of Technology, Atlanta, GA 30332, USA
          }

\date{\it Paper in refereeing stage}

\abstract{An unusual Eddington-limited thermonuclear X-ray burst was
  detected from the accreting neutron star in \bron\ with the {\it
    Rossi X-ray Timing Explorer}. The burst commenced with a brief
  (40~ms) precursor and maintained near-Eddington fluxes during the
  initial 77 s. These characteristics are indicative of a nova-like
  expulsion of a shell from the neutron star surface. Starting 122 s
  into the burst, the burst shows strong ($87\pm1$\% peak-to-peak
  amplitude) achromatic fluctuations for 60 s.  We speculate that the
  fluctuations are due to Thompson scattering by fully-ionized
  inhomogeneities in a resettling accretion disk that was disrupted by
  the effects of super-Eddington fluxes. An expanding shell may be the
  necessary prerequisite for the fluctuations.
  
}

\maketitle

\section{Introduction}
\label{intro}

Close to one hundred accreting low-magnetic-field neutron stars (NSs)
in our galaxy are known to exhibit X-ray bursts from time to
time\footnote{For a list, see
  http://www.sron.nl/$\sim$jeanz/bursterlist.html}
\citep[e.g.,][]{cor03,gal08}.  These bursts, lasting roughly 1 min and
with a k$T\approx$1 keV thermal spectrum, are due to thermonuclear
shell flashes of hydrogen and helium accreted in the top $\sim
10^8$~g~cm$^{-2}$ (or $\sim1$~m) of the NS
\citep{woo76,mar77,lam78}. The burst recurrence time depends on the
rate of fuel supply, in other words the mass accretion rate, and the
fuel mixture \citep[e.g.,][]{Fujimoto:81,lew93,bil98,stroh06}.  For
$96$\% of all 3387 cases studied by \cite{kee10}, the recurrence time
ranges from 1 hr upward. A few percent of all X-ray bursts are longer
than usual \citep[e.g.,][]{kee09}, due to larger ignition
depths. These events include ``intermediate duration'' bursts
\citep{zan05,cum06,fal08}, with durations of order half an hour, and
superbursts \citep{cor00,cum01,stroh02a}, with durations of order 1
d. These are thought to be due to thick ($\sim 10^{10}$~g~cm$^{-2}$)
helium piles in low-$\dot{M}$ systems and deep ($\sim
10^{12}$~g~cm$^{-2}$) ignition of carbon, respectively.

Helium burning through the 3$\alpha$ process is faster than hydrogen
burning. Therefore, if the fuel consists of predominantly helium, the
energy release is more likely to give rise to super-Eddington
fluxes. The resulting radiation pressure will expand the photosphere
which is easily measurable through a combination of spectral cooling
and increasing emission areas, combined with a more or less constant
bolometric flux. Photospheric radius-expansion (PRE) bursts comprise
about 20\% of all bursts \citep{gal08}. Most intermediate-duration
bursts, if not all, are of the PRE kind.

The prototype of the intermediate duration burst is from the low-mass
X-ray binary \bron\ \citep{zan05}. \bron\ is thought to be
ultracompact in nature with an orbital period of 1 hr or less, a
hydrogen-deficient donor star to the NS and a distance of 5.4 kpc
\citep{nel04,zan07}. The ultracompact nature was recently supported by
the tentative measurement of the orbital period of 17.4 min
\citep{zho10}. The burst was detected with the BeppoSAX Wide Field
Cameras and lasted approximately 40 min.  Here we report a detection
of a similar burst from the same system, but this time with the much
more sensitive Proportional Counter Array (PCA) on the Rossi X-ray
Timing Explorer (RXTE). We identify a new type of X-ray variability in
the late stages of the burst decay, which we refer to as achromatic
late-time variability. We describe the properties of this variability
and speculate on its origin.

\section{Observations} 
\label{obs}

Up to May 2010, \bron\ was observed for 520 ks with the PCA
\citep{jah06}, an instrument consisting of 5 co-aligned non-imaging
proportional counter units (PCUs) which combine to an effective area
of 6500 cm$^2$ at 6 keV, have a bandpass of 2 to 60 keV with a typical
spectral resolution of 20\% and are collimated to a field of view of
$2\degr\times2\degr$ (full-width at zero response). No other X-ray
sources are known within the field of view around \bron. Five X-ray
bursts were detected, two of which have an uncertain thermonuclear
origin \citep{gal08}. The first four occurred in 2000-2004. On Feb. 8,
2008, the PCA detected the fifth X-ray burst, when two PCUs were
active, numbers 0 and 2. The relevant RXTE observation identification
is 93416-01-05-00. The decay of the fifth burst is 7 times longer and
20\% brighter (in photon count rate per PCU) than the next
longest/brightest first burst \citep{jon01}.

\section{Analysis}
\label{ana}

\subsection{Light curves}
\label{lcs}

\begin{figure}[t]
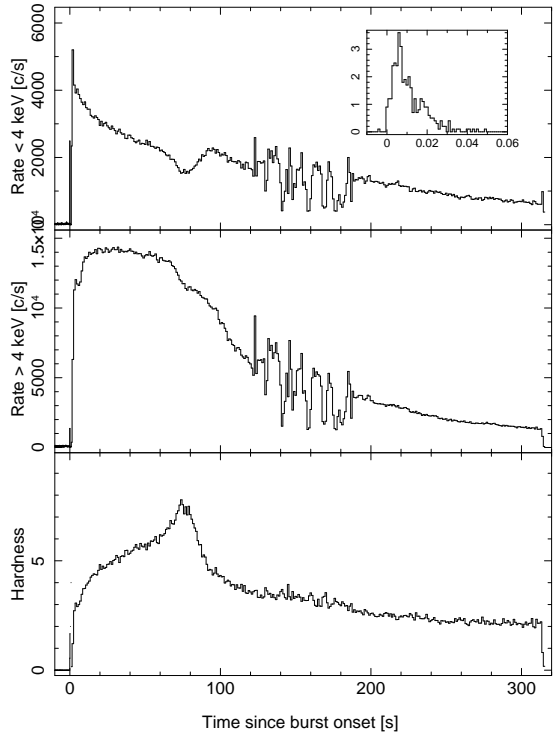

\centerline{\includegraphics[width=0.8\columnwidth,angle=0]{wholeburst.ps}}
\vspace{-9.7cm}\hspace{2.0cm}
\centerline{\includegraphics[height=0.25\columnwidth,angle=270]{precursor1.ps}}
\vspace{8cm}
\caption{PCA light curves of long burst from \bron\ in two bandpasses
  and over the 2 active PCUs (top two panels) and the hardness ratio
  between both (third panel). The time resolution is 0.1 s for times
  earlier than 1 s, and 1.0 s for later times. The inset in the first
  panel shows the full-bandpass light curve for the first 60 ms at a
  time resolution of 1 ms with rate in units of 10$^4$ c~s$^{-1}$.
\label{fig1}}
\end{figure}

\begin{figure}[t]
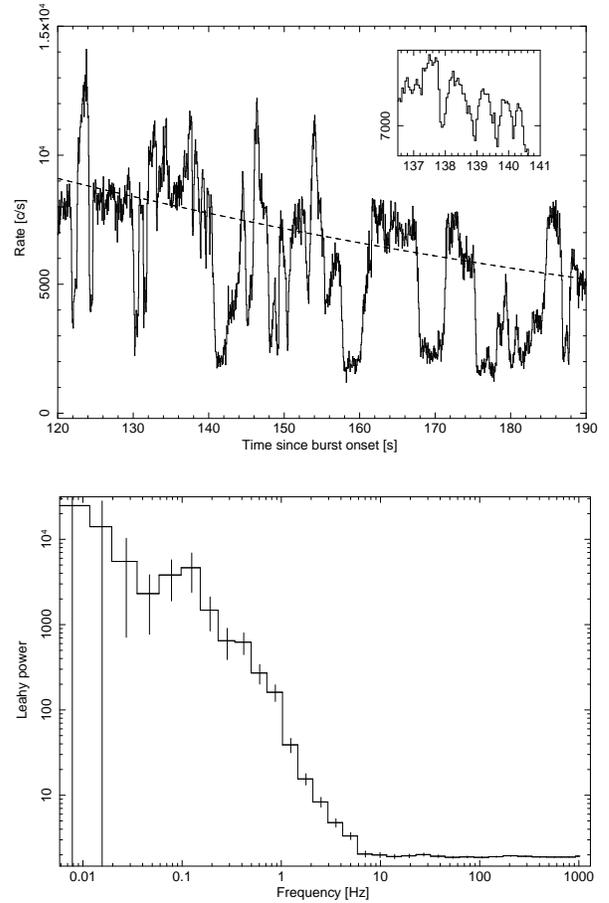

\centerline{\includegraphics[width=0.68\columnwidth,angle=270]{modulations2.ps}}
\vspace{-5.4cm}\hspace{2.0cm}
\centerline{\includegraphics[height=0.25\columnwidth,angle=270]{modulations3b.ps}}

\vspace{4.3cm}
\centerline{\includegraphics[width=0.6\columnwidth,angle=270]{power.ps}}
\caption{{\em (Top)} Full bandpass PCA light curve of a part of the
  burst that exhibits fluctuations, at 1 s resolution. The dashed line
  shows the extrapolation of an exponential function fitted to data
  beyond 190 s. The e-folding decay time fitted is
  $124.8\pm0.8$~s. The inset zooms in on the chirp. Two PCUs were
  active. {\em (Bottom)} Fourier power spectrum of same light curve,
  but at 0.5~ms resolution.
\label{fig5}}
\end{figure}

Figure~\ref{fig1} presents the light curves of the fifth burst in two
bandpasses and the (hardness) ratio of both. Many aspects of the burst
are similar to what has been observed before in this and other
sources.  The rise of the burst is very fast. In fact, a precursor can
be distinguished that merely lasts 0.04~s. This duration is similar to
that of the record holder \citep[the precursor to a burst from 4U
  0614+09;][]{kuu09}.  The start of the main event is 1.2 s after that
of the precursor. The precursor signifies that superexpansion happens
in this burst which is presumably due to an optically thick shell
being expelled by strong radiative driving \citep{zan10}. The time
scales are identical to the record burst from 4U 0614+09.  The
hardness peaks 77 s after burst onset. This identifies the so-called
'touch down' point (see spectroscopic analysis below). It coincides
with a local minimum (30\% below the trend) in the photon count rate
for energies below 4 keV. Such a decrease is hardly seen above 4 keV.

One aspect about this burst is very unusual and is the focus of the
present paper. At 122 s, a strong variability starts that lasts 66 s
with $\sim$87\% peak-to-peak amplitude.  Interestingly, the hardness
ratio is unaffected: there are no hardness fluctuations corresponding
to the intensity fluctuations. After this phase, the burst decays in
an orderly fashion until the end of the observation at 313 s when the
flux is approximately 10\% of the peak value.  The exponential decay
times during this orderly phase are $149.3\pm3.1$ and $113.4\pm1.2$~s
for the low (i.e., below 4 keV) and high-energy (above 4 keV) light
curves, respectively.

\begin{figure*}[t]
\centerline{\includegraphics[width=\columnwidth,angle=0]{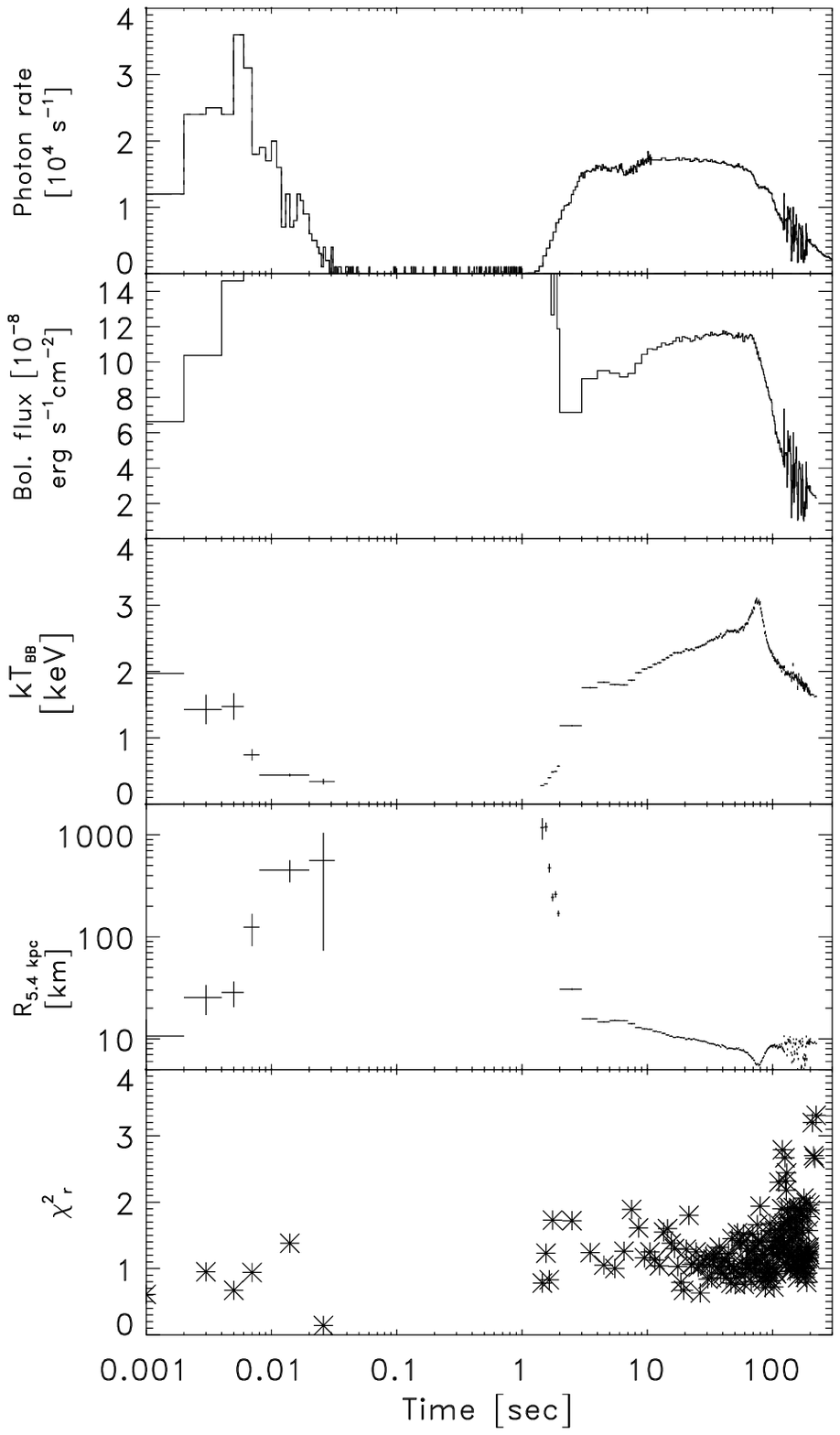}
            \includegraphics[width=\columnwidth,angle=0]{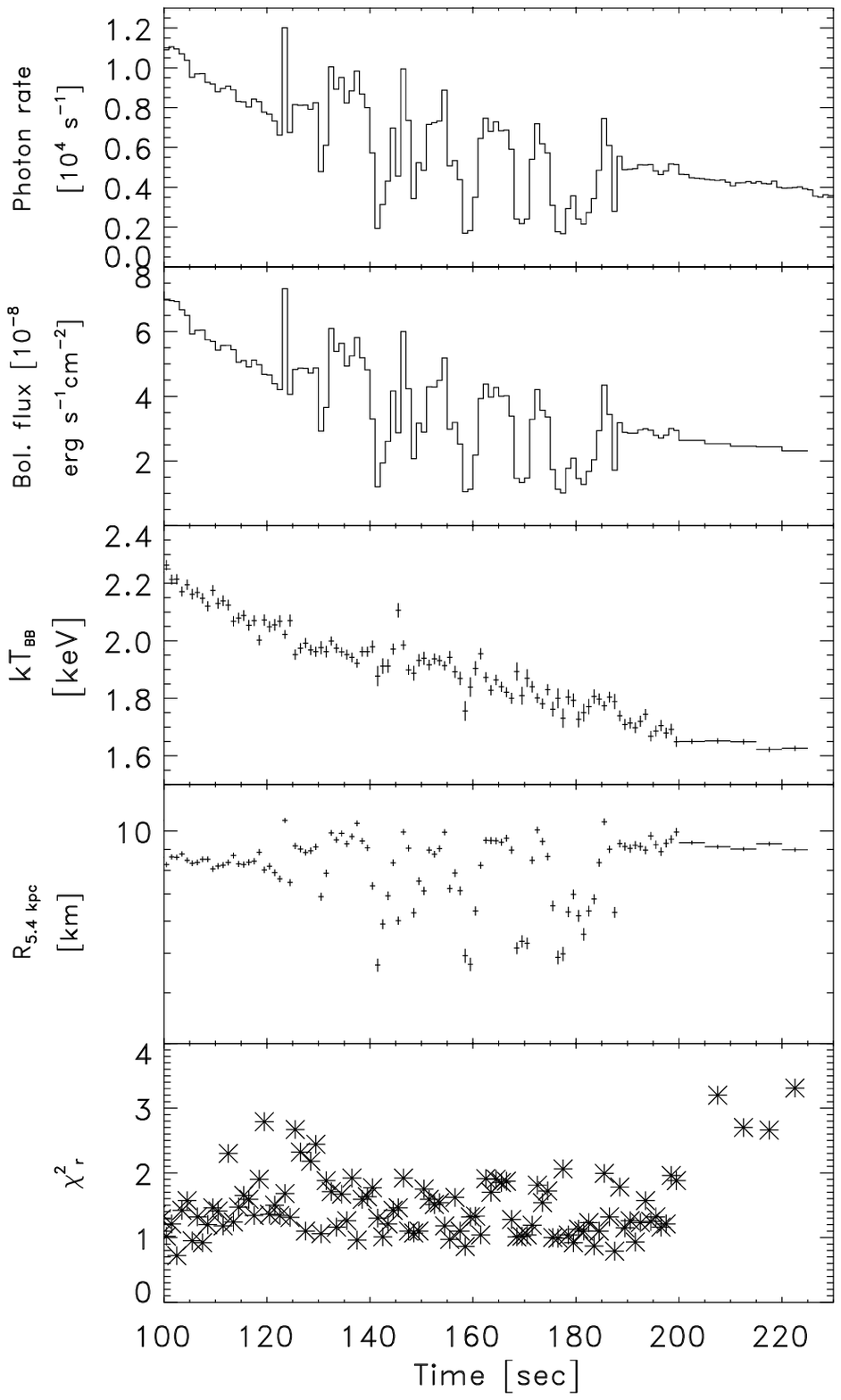}}
\caption{Time-resolved spectroscopy of the burst from \bron. The top
  panel shows the full bandpass light curve of the xenon layers, the
  2nd panel the bolometric flux (early values suffer from statistical
  errors between 10 and 20\%), the 3rd panel the black body
  temperature in terms of k$T_{\rm eff}$, the 4th panel the black body
  radius for a distance of 5.4 kpc, and the bottom panel the reduced
  $\chi^2$ for the fits per time bin.
\label{fig3}}
\end{figure*}

Figure~\ref{fig5} shows a detailed light curve of the variable phase
and a Fourier power spectrum of the same data.  A number of features
to note here are:
\begin{list}{\leftmargin=0.4cm \itemsep=0cm \parsep=0cm \topsep=0cm}
\item
\item[$\bullet$] the typical time scale is 1~s. There is no
  variability below 0.2~s; the Fourier power density above 5 Hz is
  consistent with statistical (white) noise with an upper limit on the
  fractional rms of about 1.2\% at the frequency resolution applied in
  Fig.~\ref{fig5};
\item[$\bullet$] A cross correlation analysis between the low
  ($<4$~keV) and high energy ($>4$~keV) lightcurves shows no time lags
  in excess of 0.01~s;
\item[$\bullet$] No burst oscillations were detected in any time
  segment, with 3$\sigma$ upper limits on the fractional rms amplitude
  \citep[c.f.,][]{stroh06} of 3.9\% in 4~s time segments and 6.9\% in
  1~s segments;
\item[$\bullet$] most dip ingresses are shorter than a few tenths of a
  second.  Egresses are almost always longer;
\item[$\bullet$] dip durations range from 0.2 to 9 s. Dips tend to
  become longer at later times;
\item[$\bullet$] fluctuations upward are limited to a factor of 1.6
  with respect to the general decay trend (dashed line in
  Fig.~\ref{fig5}). Fluctuations downward are down to a factor of 3.
  The peak-to-peak amplitude for the whole period of fluctuations is
  $87\pm1$\%;
\item[$\bullet$] all upward fluctuations are left and right
  accompanied by downward fluctuations, most clearly for the features
  at 124 and 134 s. The converse does not apply;
\item[$\bullet$] between 138 and 140 s, 4 dips at progressively
  shorter intervals are seen, giving the appearance of a chirp.
\end{list}

\subsection{Spectra}
\label{spec}

We employed event mode data that have a time resolution of 125 $\mu$s
and bin the photon energy in 64 channels. Data from all Xenon detector
layers and the two active PCUs were combined. Spectra were extracted
in time bins that are sufficiently small to follow the variations in
hardness (see Fig.~\ref{fig1}), with bin sizes measuring 2 ms (until 6
ms after precursor onset; see Fig.~\ref{fig3}), 12 ms (until 1 s), 0.1
s (until 2 s), 1 s (until 200 s) and 5 s (beyond 200 s). The particle
and cosmic background were modeled per time bin as prescribed by {\tt
  pcabackest} with the minimum allowed sample time of 16~s.  A
response matrix was calculated through {\tt pcarmf} version
11.7. Exposure times were corrected for detector dead time following
RXTE Cook
Book\footnote{http://heasarc.gsfc.nasa.gov/docs/xte/recipes/pca\_deadtime.html}.
Since the time resolution for such a correction is limited to 0.125 s,
interpolations were necessary for the precursor. The spectra were
analyzed between 3 and 30 keV with {\em XSPEC} version 12.5.0
\citep{arn96}. A systematic error of 0.5\% was assumed per photon
energy channel. All spectra were corrected for the modeled particle
and cosmic background.

A pre-burst spectrum of \bron\ was acquired from 672 s of standard-2
data that are available up to 100 s before the burst onset. The shape
of the pre-burst spectrum is within 3 to 30 keV consistent with the
broadband 0.1-200 keV spectrum that was measured by \cite{zan05}. The
extrapolated 0.1-200 keV unabsorbed flux is 50\% larger at
$5.5\times10^{-10}$~\ecs\ than the absorbed 3 to 30 keV flux. This is
0.5\% of the bolometric peak flux of the burst (see below). For a
canonical 1.4~M$_\odot$/R=10~km NS, this translates to a mass
accretion rate of $1.0\times10^{16}$~g~s$^{-1}$. Allowance was made
for the accretion radiation in the spectral analysis of the burst data
by including a fixed pre-burst model, thus indirectly subtracting the
background. This was not done for the precursor data since it is
expected that the accretion flow is significantly disturbed (i.e., the
flux during superexpansion is lower than before the
precursor). Although the accretion radiation was accounted for, it was
found that neglecting this did not make a significant difference for
any of the spectral results. The accretion flux is always much lower
than the burst flux for data with a meaningful statistical
significance.

The time-resolved burst spectra were modeled by simple black body
radiation with interstellar absorption. The absorbing column was fixed
at $N_{\rm H}=3\times 10^{21}$~cm$^{-2}$ \citep{zan05}, which is
hardly detectable with the 3-20 keV bandpass of the PCA. The results
are shown in Fig.~\ref{fig3}, in both logarithmic (left panel; to
focus on the early part of the burst) and linear scales (right panel;
to focus on the part where the fluctuations occur). The bolometric
fluence over the measurements of the burst is
$(1.58\pm0.11)\times10^{-5}$~erg~cm$^{-2}$. Extrapolating the decay
beyond the end of the observation (through an exponential function
with an e-folding decay time of 120~s) predicts that 20\% of fluence
is likely to have been missed in the observation. For 5.4 kpc, this
translates to a total radiative energy output of about 7$\times
10^{40}$~erg, which is consistent with an ignition depth of $4\times
10^{9}$~g~cm$^{-2}$ if the pre-burst layer is pure helium \cite[see
  Fig.~8 in][]{zan05}.

The precursor shows a rapid photospheric expansion driven by radiation
pressure as the burst emission is super-Eddington. Within 30 ms, the
photosphere expands to roughly $10^3$~km, with an average velocity of
roughly $0.1c$. We do not believe this is an optical depth effect,
because prior to the burst no gas of sufficient density is expected up
to that height around the NS. It would fall to the NS within a
dynamical time scale of 1~ms. The measured average velocity is roughly
20\% of the escape velocity from a canonical NS, so this shows that
the photosphere is strongly driven by radiation for as long as it is
visibly expanding. Otherwise it would return to the NS already within
a height of less than 1 km.  The photosphere is probably on a shell of
thickness $10^{6-7}$~g~cm$^{-2}$ \citep[c.f.,][]{zan10}. At the end of
the precursor the shell radiation moves out of the PCA bandpass so
that it becomes undetectable. After 1.2~s the burst emission
reappears. Probably the shell at that time diffused so much that it
became optically thin and uncovered the underlying radiating NS. The
radiation is still super-Eddington with high temperatures ($>$2 keV),
moderately increased levels of black body radius and a more or less
constant bolometric flux. After 70 s the flux starts to decrease,
signifying the end of the Eddington-limited phase. At that time the
spectrum shows a hardening lasting approximately 40 s which may be due
to Comptonization of the thermal photons against a hot receding
atmosphere \citep{lon86,Madej:04}. The simultaneous radius
measurements are most probably biased by this systematic effect and do
not represent true variations of the photospheric size. This
'touch-down' phase is often seen in Eddington-limited bursts
\citep{dam90,gal08,stei10}. Subsequently ordinary NS cooling sets in,
until the fluctuations start at 120 s. As is the case for the hardness
ratio in Fig.~\ref{fig1}, the evolution of the black-body temperature
is largely unaffected by the fluctuations. The root mean square of the
temperature variations relative to a fitted exponential function is
$1.0\pm1.6$\% prior to the fluctuations (i.e., between 80 and 120 s)
and $2.3\pm2.3$\% during (i.e., between 120 and 190 s). We conclude
that the fluctuations are achromatic.

\begin{figure}[t]
\centerline{\includegraphics[width=0.8\columnwidth,angle=0]{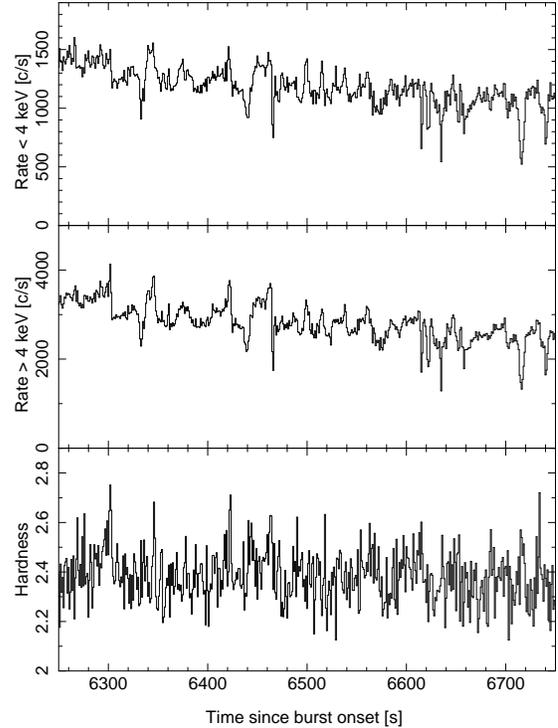}}
\caption{PCA light curve of the part of the superburst from 4U 1820-30
  that shows achromatic variability. For panel explanation, see
  Fig~.\ref{fig1}. The time resolution is 1.0 s throughout.
\label{fig4}}
\end{figure}

\section{Achromatic late-time variability in other bursts}
\label{sec:others}

Achromatic late-time variability was also seen in the previous long
burst detected from \bron, with BeppoSAX-WFC \citep[][referred to in
  \S~1]{zan05}. In fact, it looks rather similar, but with worse
statistical quality (the effective area of BeppoSAX-WFC is 45 times
smaller than of the PCA). The fluctuations started 135~s after burst
onset, lasted 50 s and the amplitudes were factors up to
3. Additionally, a very long dip was detected starting 200 s after
burst onset. The flux decreased by a factor of 4 for 70 s. The burst
decay is followed {\em during the dip}. Also for this dip no spectral
change could be detected.

The WFC burst was probably also a superexpansion burst, because it
shows a rapid decline in black body radius at the onset. A definitive
identification of superexpansion is not possible since a precursor is
not detected, but this is not unexpected because there would be too
few photons in the precursor to be able to detect it with the
moderately sensitive WFC. The WFC burst has, within 20\%, the same
near-Eddington duration and decay time as the PCA burst. These
similarities are unlikely to be a coincidence and point to the burst
as the initiator of the variability.

Achromatic variability was also detected in two other long
superexpansion bursts, from SLX 1735-269 \citep{mol05} and 4U 1820-30
\citep{stroh02a,bal04,zan10}. In the former, detected with JEM-X on
INTEGRAL, the fluctuations lasted between 700 and 1300 s after burst
onset, with a touch down at 400 s.  The burst from 4U 1820-30 was a
superburst detected with the PCA. Fluctuations in that burst have
varying time scales, but those at $\sim10$~s time scale are also
energy independent (see Fig.~\ref{fig4}). The fluctuations started
3365 s after burst onset and lasted until 7200 s, while the touch down
for that burst is at 1100 s.

Finally, achromatic variability was seen in a burst from M15 X-2
\citep{jvp90} between 127 and 156 s after burst onset, with a touch
down time of 94 s. They were attributed to dipping activity, although
it was not understood why the non-burst emission failed to exhibit
dips.

Achromatic variability was not seen in shorter bursts from \bron\ that
are Eddington-limited but lack superexpansion \citep{jon01,zan05}.

Fluctuations downward were seen in a burst from A~1246-58
\citep{zan08}. A number of intermediate duration bursts which (are
likely to) have a precursor did not show fluctuations
\citep[e.g.,][]{zan07,zan10}.

In conclusion, based on the aforementioned 6 fluctuating bursts, the
necessary conditions for fluctuations seem to be the presence of
superexpansion (i.e., the expulsion of a shell), a burst duration in
excess of 120~s (although that is susceptible to selection effects),
and the termination of the Eddington-limited phase. Since some long
and probably superexpansion bursts lack fluctuations, another
parameter may be at work, perhaps the inclination angle. Since none of
the above bursters show dipping or eclipsing behavior in the
persistent emission, the inclination angle must be smaller than
$i=70$\degr\ \citep{whi95}. It seems plausible that in bursters which
fail to show fluctuations in superexpansion bursts, the inclination
angle is smaller again, so that a disrupted disk does not extend into
the line of sight.

\section{Discussion}

We presented an analysis of a type of X-ray intensity variation seen
in some long intense X-ray bursts with superexpansion. Such variations
have not been reported previously in such detail; here we discuss
possible origins.

\subsection{Origin of fluctuations}

The fluctuations are achromatic and go factors above and below the
decay trend of the burst. The latter rules out a thermonuclear origin,
since then the flux would not undercut the decaying trend of the
initial flash. An example of thermonuclear variability is millihertz
oscillatory behavior in the non-burst emission of some bursters
\citep[4U 1608-52 and 4U 1636-536;][]{rev01,heg07b,alt08}. The
characteristics of these oscillations are different from the
achromatic variability reported here, both in time scales and spectral
behavior.

The fluctuations downward point to something occasionally and partly
obscuring the line of sight to the NS. Since the NS is small and the
dips are $\approx70$\% deep, the obscuration appears to be due to
clouds of optical depth $\tau\approx1$.  The fact that the
fluctuations are achromatic suggests that the cloud plasma is in
thermal balance with the radiation field so that the scattering of the
photons is elastic (i.e., Thompson scattering). The column density of
the electrons in the clouds, for a uniform density, would be $N_{\rm
  e}=\tau/\sigma_{\rm T}=1.7\times 10^{24}$~cm$^{-2}$ (with
$\sigma_{\rm T}$ the Thompson cross section). Since the clouds need to
be larger than the $L=$20 km size of the NS, the electron density is
inferred to be $n_{\rm e}<N_{\rm e}/L\approx 10^{18}$~cm$^{-3}$.

The picture arises that hot plasma clouds move around the NS and
Thompson scatter the burst photons. If a cloud is in front of the NS,
burst emission is scattered out of the line of sight. If it is at the
other side of the NS, but visible to the observer, it is scattered
backward into the line of sight. Since between upward and downward
fluctuations we see basically unaffected burst emission (for instance
at 125-129~s, 134-136~s, 162-166~s) only a few clouds would be
responsible for the fluctuations and there would hardly be any sideways
scattering. The latter would be consistent with Thompson scattering,
since that is predominantly in the forward and backward
direction. Still, the flux level between fluctuations appears to be
somewhat increased. This may be due to some sideways scattering or to
backward scattering of clouds with smaller angular sizes.

The fluctuations downward are reminiscent of dips as seen in the
out-of-burst radiation of low-mass X-ray binaries with inclination
angles higher than about 70$\degr$. Those are thought to be due to
splashes off the accretion disk that result from the impact of the
mass stream from the donor star on the disk \citep{par86,arm98}. The
splashes usually populate a small range of orbital phases and return
to the line of sight periodically, obscuring the inner (X-ray
radiating) disk, resulting in dips in the light curve with typical
durations of a few tens of seconds. The dip spectrum can be modeled by
a combination of energy-dependent absorption by cold matter and
energy-independent absorption by warm photoionized matter with a
photo-ionization parameter of $\xi=L/n_{\rm e}r^2\sim10^{2.5-3}$
erg~s$^{-1}$cm \citep{boi05,dia06}. The dips reported here are not
connected to certain orbital phases, but to certain phases in long
bursts. A simple estimate for the probability that the two long bursts
occur at the same orbital phase by chance is 1~min/17.4~min which is
only 6\%. Also, classical dips in the non-bursting emission would also
have to occur, but have never been reported for \bron. Still, our dips
are energy-independent like part of the classical dips.  For
log$\xi\ga 3$, $L\approx 2\times10^{38}$~\lum\ and $n_{\rm e}\la
10^{18}$~cm$^{-3}$, the distance of the scattering medium to the NS
would be of order $r\la 10^4$~km.

The first down-up-down fluctuation at 124 s appears to be the effect
of a single cloud that first is in front of the NS, 1~s later behind,
then again in front and subsequently gone, possibly accreted. Almost
the same happens at 130~s. This suggests a 2~s Keplerian period, which
would translate to a typical cloud distance from the NS of $3\times
10^3$~km which is roughly consistent with the photo-ionizing argument
above. The chirp at 139~s could be explained similarly, with a
progressively shorter orbital period. Later dips appear to become
progressively longer, but peaks stay as long and disappear. In the
other superexpansion burst detected from \bron\ by the WFC, one dip
lasted as long as 70~s. An explanation may be that the clouds extend
over larger azimuthal ranges as time progresses and obscure
backscattering. Either the clouds become larger in linear size, or
they remain as large but come closer to the NS.

Warping of accretion disks may be strengthened by irradiation from a
central source. Previous studies \citep{pet77,pri96,pri97,wij99,ogi01}
assumed the central source to be the inner accretion disk, but a
bursting neutron star would seem viable as well. However, the time
scales of the radiation are totally different (weeks versus minutes)
and it is questionable whether burst radiation will be as effective in
warping the disk. Furthermore, radiative warping would occur
irrespective of the presence of superexpansion, in contrast to what we
observe.

\subsection{Accretion disk disruption}

Where would the cloud structure come from?  An obvious possibility is
that the clouds arise from material in the inner accretion
disk. Theoretical investigations have shown that it is possible for
Eddington-limited X-ray bursts to disrupt the inner accretion disk.
\cite{bal05} treated this issue most recently in an attempt to explain
spectroscopic measurements of the 4U 1820-30 superburst (not the
aforementioned fluctuations); earlier discussions may be found in
\cite{fuk83}, \cite{wal89}, \cite{wal92} and \cite{fuk95}. Five
mechanisms are discussed by \cite{bal05}, four of which are concerned
with the interaction between the radiation from the NS with the
accretion disk: radiatively and thermally driven winds, increased
inflows as a result of radiation drag (i.e., the Poynting-Robertson
effect), and X-ray heating of the disk. It is clear that thermally
driven winds are ineffective due to the large gravitation field near
the NS. Radiation pressure is more effective, but it is not obvious
that it can sweep the inner accretion because the optical depth of the
disk is large along the radial travel direction of the burst
radiation.  Radiation drag is probably only important in the innermost
region of the accretion disk \citep[c.f.,][]{mil96}, at radii smaller
than 20$R_{\rm g}\approx40$~km with $R_{\rm g}=GM/c^2$. The final
radiative disruption mechanism, X-ray heating of the disk surface, may
change the disk structure through for example puffing up the surface
on a dynamical time scale and settling it on viscous time scales. A
non-radiative fifth mechanism is concerned with the mechanical
interaction between the radiatively driven outflow and the accretion
disk. This effect was dismissed by \cite{bal05} for a radiation-driven
wind on the basis of that carrying insufficient kinetic energy to
overcome the potential energy of the wind. However, the circumstances
may be different for superexpansion bursts where the shell carries
much more density than the previously considered disk. 
In \S \ref{sec:disshell} and \ref{sec:disrad} we concentrate on
the two most likely mechanisms for disrupting the disk.

\subsubsection{Disruption by shell}
\label{sec:disshell}

The geometrically thin but optically thick shell expands at a velocity
of order 10\% of the speed of light. The column thickness of the shell
at launch is at maximum $\sim$1\% of the ignition depth or $y \la
10^{6-7}$~g~cm$^{-2}$ and diffuses as $R^{-2}$. The radial profile of
the density of the accretion disk at mid plane, for a gas-supported
disk (as applicable for an accretion rate of $\sim$1\% of Eddington)
with electron-scattering opacity \citep{ss73,sve94,bal05} and assuming
a Gaussian vertical density profile, is given by
\begin{eqnarray}
\rho & = & \rho_0~(R/R_{\rm g})^{-33/20}~J(R)^{2/5}{\rm g~cm}^{-3}
\label{eq:density}
\end{eqnarray}
with $R_{\rm g} = 2 M_{1.4}~{\rm km}$ ($M_{1.4}$ is the NS mass in units of
1.4~M$_\odot$),
\begin{eqnarray}
J(R) & = & 1 - \sqrt{6R_{\rm g}/R}
\end{eqnarray}
and
\begin{eqnarray}
\rho_0 & = & 47~d_{5.4}^{6/5}\dot{M_{16}}^{2/5}\alpha_{0.1}^{-7/10} M_{1.4}^{-11/10}~{\rm g~cm}^{-3}
\end{eqnarray}
($d_{5.4}$ is the distance in units of 5.4 kpc, $\dot{M_{16}}$ the
accretion rate in units of $10^{16}$~g~s$^{-1}$ and $\alpha_{0.1}$ the
disk viscosity parameter in units of 0.1).  The mean molecular weight
is assumed to be $\mu=1.3$. For $\alpha=0.1$, $\dot{M}_{16}$ between 1
and 10 (for radiation efficiencies of $\eta=1$ and 0.1, respectively)
and $y=10^6$ g~cm$^{-2}$, the mass column swept up by the diffusing
shell will have reached the shell's own diminished column mass at 50
to 150 km above the NS surface. For $y=10^7$ g~cm$^{-2}$ this range
becomes 110-440 km. We assume here that the disk is magnetically
truncated at $25 \dot{M}_{16}^{-2/7}$ km from the NS \citep[for a
  canonical NS with $B=10^8$~G; this does not have to be accompanied
  by pulsations, see][]{fra02}. Away from the mid plane the disk
density is lower.  At 3-4 scale heights above or below the disk (this
is approximately equivalent to the $\tau=1$ level), the shell may be
able to disrupt the disk up to a distance of $\sim3000$~km. Thus, the
shell appears able to blast away a small radial part of the disk, out
to perhaps a few tens of NS radii, and ablate the surface of the disk
up to distances of $10^{3-4}$~km. Note that with a suggested orbital
period of 17.4~min~\citep{zho10}, the accretion disk has a size of
$1.7\times 10^5$~km.

It is difficult to determine whether the energy and momentum are
conserved in the the disk/shell interaction. The apparent expansion
velocity of the shell is only $\sim$20\% of the escape velocity at the
NS surface which allows to bridge only 4\% of the potential energy
required to escape, while the shell is observed to bridge 99\% (i.e.,
the maximum photospheric radius is $10^3$~km). Obviously, the
transformation of radiation to kinetic energy is a dominant term in
the energy balance. An assessment of that is beyond the scope of the
present observational paper. Conservation of momentum is roughly
upheld.  The ram pressure of the shell at 100 km is $\rho$v$^2=y
(R/R_{\rm NS})^2 (0.1c)^2/D\approx10^{17}$~dyne~cm$^{-2}$ with $D=1$~m
the thickness of the shell or perhaps a few orders of magnitude
smaller if the shell thickness grows with the shell radius.  The
density of the disk at 100 km is of order 10$^{-3}$~g~cm$^{-3}$
(Eq. \ref{eq:density}). From $\dot{M}=2\pi R H\rho v$ this results in
a radial inward velocity of $\approx20$~km~s$^{-1}$. The ram pressure
of the disk\footnote{It is noted that the larger disk ram pressures
  calculated in \cite{fra02} are for azimuthal motion instead of
  radial, as is applicable for balancing magnetic pressure} then is
$\approx10^{7}$~dyne~cm$^{-2}$ at 100 km. However, the gas pressure of
the disk is larger and will take over counteracting the shell. For an
ideal gas it is $\sim10^{15}$~dyne~cm$^{-2}$ for a disk temperature of
10$^7$~K. This order-of-magnitude calculation shows that the shell may
have enough momentum to sweep the inner 100 km of the accretion disk.

\subsubsection{Disruption by radiation pressure}
\label{sec:disrad}

In the previous section it is shown that in the first 10$^{-1}$~s the
shell may disrupt the accretion disk over the inner $10^{3-4}$~km,
possibly even clearing it completely over the inner $10^2$~km. Thus,
an environment is created which is more susceptible to the
near-Eddington fluxes that ensue for 70~s. \cite{bal05} modeled the
effects of super-Eddington fluxes on the accretion disk without the
preceding shell disruption, but with otherwise similar circumstances
(same burst flux that stays near-Eddington for at least 70 s; same
type of binary orbit). They find that the gas in the surface layers of
the disk with $n_{\rm e}=10^{17}$~cm$^{-3}$ is accelerated radially to
$\approx$0.1~c within at most $\approx30$~s. Thus, the radiation is
likely to further clear the parts of the disk already disrupted by the
shell or, at least, prevent it from falling back. The effect may even
be larger, because the shell will likely puff up the disk, increasing
the solid angle of accretion disk material as seen from the NS so that
a larger portion of the burst luminosity will be involved in the
radiative disruption.

\subsection{Accretion disk resettlement}

If the disk is disrupted, it is plausible that it deviates
considerably from cylindrical symmetry. The push by the shell and its
lateral expansion will likely puff up the disk, as will the radiation
pressure due to turbulence. If the effect of super-Eddington radiation
subsides, the accretion disk is expected to return to the immediate
surroundings of the NS, because the feeding of the disk by the donor
star continues. We speculate that it is this return of the disk that
is responsible for the achromatic fluctuations.

How can, then, the delay time be explained of 45 s between the end of
the Eddington-limited phase and the onset of achromatic variability?
The free fall time over 10$^3$~km of 0.5 s is much shorter, so it
involves something else than a straightforward fall back of disk
material. Could it be the viscous dissipation time scale of the disk?
This time scale is, for a gas-pressure dominated accretion disk, given
by \citep{sve94,bal05}
\begin{eqnarray}
t_{\rm visc} & = & 4 \alpha_{0.1}^{-4/5} M_{1.4}^{8/5}
\dot{M}_{16}^{-2/5} (R/R_{\rm g})^{7/5} J(R)^{-2/5}~~{\rm s}.
\end{eqnarray}
At 100 km, $t_{\rm visc}$ ranges between 60 and 960 s. This is in line
with the measured delay time of 45 s. However, not so for the 2000 s
delay time for the superburst from 4U 1820-30. $\dot{M}_{16}$ is 11
times larger and $t_{\rm visc}$ 2.5 times smaller while the measured
value is 50 times larger.

Rather than the time it takes the disk to respond to changes in the
environment, the onset of the fluctuations may be determined by the
level of irradiation (see \S~\ref{sec:disrad}).  The radiation
pressure is equal to $L/4\pi R^2c$. The disk will be disrupted as long
as $L>4\pi R^2 c P_{\rm ram}$ with $P_{\rm ram}$ the (uncertain) ram
pressure of the in-flowing accretion disk. For $L=2\times 10^{38}
d_{5.4}^2$~\lum\ and $R=100$~km, $P_{\rm ram}$ would have to be
$5\times 10^{13}$~dyne~cm$^{-2}$. The uncertainties are large in the
disk parameters ($\dot{M}$, $T$, $R$, $P_{\rm ram}$) and the physics
employed may be oversimplified. Therefore, it is difficult to judge
the validity of this effect.

The duration of the achromatic variability, 66 s, may be related to
the radial extent over which the clouds reside when they start to fall
back. The longer the near-Eddington phase lasts, the further away disk
material is blown off by the radiation pressure. Indeed, there is a
relation between the duration of the variability and that of
near-Eddington fluxes of the cases discussed in \S~\ref{sec:others}
(roughly linear).

\section{Summary and outlook}

We report achromatic late-time variability in thermonuclear X-ray
bursts from \bron\ and a few other low-mass X-ray binaries, and
suggest it may be explained by the settlement of the accretion disk
near the NS that is disrupted by the combination of a shell expulsion
and a near-Eddington flux.  Order-of-magnitude calculations of the
effect of the various disruption mechanisms suggest that this
explanation is plausible, although there is a need for follow up by
numerical time-dependent calculations of the dynamical response of the
accretion disk to the expanding shell and wind, the near-Eddington
radiation and the subsequent settling disk. We note that such a
numerical investigation may also be interesting for the study of less
energetic classical novae on massive white dwarfs where the shell is
expected to be less massive so that the disk may survive the explosion
\citep{sok08}. Besides these theoretical investigations, detecting
more bursts with this variability with a variety of burst durations
would be constraining, particularly from X-ray binaries with known
orbital periods. An observation that would reject our hypothesis would
be achromatic variability in bursts without superexpansion with a
device that is sensitive enough to detect $10$~ms precursors. Finally,
we note that our findings imply a warning: not all features in X-ray
burst light curves originate on the neutron star \citep[see
  also][]{sha03} and in some cases these additional effects can be
significant.

\acknowledgements

We sincerely thank Jean-Pierre Lasota, Laurens Keek, Dimitrios
Psaltis, Jennifer Sokoloski and Nevin Weinberg for useful discussions.

\bibliographystyle{aa} \bibliography{references}

\end{document}